\documentclass[prl,reprint,amsmath,amssymb,aps,preprintnumbers,showpacs]{revtex4-1}
\usepackage[colorlinks=true,linkcolor=black, citecolor=black,urlcolor=black]{hyperref} 
\usepackage{amstext,amsmath,amssymb,amsfonts,mathrsfs}   
\usepackage{graphicx}  
\usepackage{braket} 
\usepackage{bbm} 
\usepackage[american,USenglish,british,UKenglish]{babel}
\usepackage[dvipsnames]{xcolor}
 \usepackage{tikz}
\usepackage{pgfplots}
\usepgfplotslibrary{external}

\pgfplotsset{compat=1.15}

\newcommand{\AdS}{\text{AdS}}
\renewcommand{\S}{\text{S}}
\newcommand{\T}{\text{T}}

\newcommand{\alg}[1]{\text{#1}}

\renewcommand{\sl}{\alg{sl}}

\newcommand{\su}{\alg{su}}
\newcommand{\psu}{\alg{psu}}

\newcommand{\sL}{\mbox{\tiny L}}
\newcommand{\sR}{\mbox{\tiny R}}

\newcommand{\acomm}[2]{\big\{#1,\,#2\big\}}

\newcommand{\Ql}[1]{\mathbf{Q}^{#1}}
\newcommand{\Qr}[1]{\widetilde{\mathbf{Q}}_{#1}}
\newcommand{\Sl}[1]{\mathbf{S}_{#1}}
\newcommand{\Sr}[1]{\widetilde{\mathbf{S}}^{#1}}
\newcommand{\Hl}{\mathbf{h}}
\newcommand{\Hr}{\widetilde{\mathbf{h}}}
\newcommand{\Htot}{\mathbf{H}}
\newcommand{\M}{\mathbf{M}}
\newcommand{\C}{\mathbf{c}}

\begin{document}

\vspace*{0cm}

\title{Long Strings and Symmetric Product Orbifold from the AdS\texorpdfstring{$_{\boldsymbol 3}$}{3} Bethe Equations}
\author{Alessandro Sfondrini}
\email{alessandro.sfondrini@unipd.it}

\affiliation{Dipartimento di Fisica e Astronomia, Universit\`a degli Studi di Padova, via Marzolo 8, 35131 Padova, Italy\\
Istituto Nazionale di Fisica e Nucleare, Sezione di Padova, via Marzolo 8, 35131 Padova, Italy}
\date{\today}

\begin{abstract}
\noindent
A particularly rich class of integrable systems arises from the AdS/CFT duality. There, the two-dimensional quantum field theory living on the string worldsheet may be understood in terms of a non-relativistic factorized S matrix, and the energy spectrum may be derived by techniques such as the mirror thermodynamic Bethe ansatz or the quantum spectral curve. In the case of AdS$_{3}$/CFT$_2$ without Ramond-Ramond fluxes, the worldhseet theory is a Wess-Zumino-Witten model with continous and discrete representations which, for the lowest allowed level, is dual to the symmetric product orbifold of a free theory. I will show how  continuous representations may arise from integrability, and that at lowest level the Bethe equations yield the symmetric product orbifold partition function on the nose.
\end{abstract}

\maketitle

\paragraph{Introduction.}
Exactly-solvable models play a crucial role in theoretical physics. 
In the quantum world, they arise in low dimension: Integrable spin-chains, lattice models, two-dimensional conformal field theories (CFTs) and integrable quantum field theories (IQFTs). Such  models have \textit{infinite-dimensional symmetries} which characterise them entirely. A prototypical century-old example is the Heisenberg spin chain solved by Bethe~\cite{Bethe:1931hc}. The Bethe ansatz, as it is now known, assumes that wavefunctions are given by a superposition of plane waves, characterised by a discrete set of excitation numbers, and that the energy is an additive functional thereof. Over the years, the approach was put on firmer ground (the \textit{algebraic} Bethe ansatz, see \textit{e.g.}\ Ref. \cite{Faddeev:1996iy}) and extended to other setups, most strikingly to IQFTs, see \textit{e.g.}\ Refs.~\cite{Dorey:1996gd,Bombardelli:2016rwb} for an introduction. While strictly speaking QFTs do not admit wavefunctions, IQFTs feature no macroscopic particle production and this, together with the factorization of their S~matrix, makes them tractable~\cite{Zamolodchikov:1978xm}. Still, the Bethe ansatz is only approximate in that case, due to finite-volume ``wrapping'' effects~\cite{Luscher:1985dn,Luscher:1986pf}. These can be accounted for by exploiting the thermodynamic Bethe ansatz~\cite{Zamolodchikov:1989cf,Dorey:1996re}, originally developed to study \textit{finite-temperature} rather than \textit{finite-volume} physics~\cite{Yang:1968rm}. (As one needs to exchange space and time, this is often called the \textit{mirror} thermodynamic Bethe ansatz, mTBA~\cite{Arutyunov:2007tc}.) 
A somewhat orthogonal approach to exact solvability arises in CFT, see \textit{e.g.}\ Ref.~\cite{DiFrancesco:1997nk}: There, one does not consider particles, but instead focuses on the representations of the Virasoro algebra and of other current algebras. In some cases, this is enough to  construct the whole spectrum of the theory and possibly its correlators. Physically, one would expect these two approaches to be related as IQFTs may arise from perturbing a CFT by a relevant (or irrelevant~\cite{Smirnov:2016lqw,Cavaglia:2016oda}) operator. Mathematically the relationship is not straightforward and still not completely understood, see \textit{e.g.}\ Refs.~\cite{Bazhanov:1994ft,Bazhanov:1996dr, Bazhanov:1996aq,Bazhanov:1998dq,Dorey:2007zx,Dorey:2019ngq}.

The aim of this letter is to explore and relate various exactly-solvable models arising from a special instance of the AdS$_{d+1}$/CFT$_{d}$ correspondence~\cite{Maldacena:1997re,Witten:1998qj,Gubser:1998bc} between strings on $(d+1)$-dimensional anti-De Sitter space and $d$-dimensional CFTs. It is by now well-known that rich non-relativistic integrable models may live on the two-dimensional string worldsheet (and equivalently in a large-$N_c$ limit of the dual CFT~\cite{tHooft:1973alw}), see Refs.~\cite{Arutyunov:2009ga,Beisert:2010jr} for reviews.
I will focus to $d=2$, where the dual CFT is also two-dimensional. As it turns out~\cite{Maldacena:1997re}, the string theory may feature a mixture of Neveu-Schwarz-Neveu-Schwarz (NSNS) and Ramond-Ramond (RR) fluxes. All these setups are classically integrable~\cite{Babichenko:2009dk, Cagnazzo:2012se} and can support a factorized integrable S~matrix~\cite{Borsato:2012ud,Hoare:2013pma,Lloyd:2014bsa,Borsato:2014exa,Lloyd:2014bsa,Borsato:2015mma}, see Ref.~\cite{Sfondrini:2014via} for a review. Here I will mostly restrict to the case of pure-NSNS fluxes. Here, and only here, a simple \textit{worldsheet CFT} description also exists, featuring a (supersymmetric) $\sl(2,\mathbb{R})_{k}$ Wess-Zumino-Witten model~\cite{Maldacena:2000hw,Maldacena:2000kv,Maldacena:2001km} in the Ramond-Neveu-Schwarz formalism. Correspondingly, the integrable structure drastically simplifies and the mTBA can be solved in closed form~\cite{Baggio:2018gct,Dei:2018mfl}.
Furthermore, for the lowest allowed level $k=1$, the dual CFT$_2$ is also known~\cite{Giribet:2018ada,Gaberdiel:2018rqv, Eberhardt:2018ouy}: It is the symmetric-product orbifold CFT of $\T^4$, $\text{Sym}^{N_c}T^4$, obtained from quotienting by the symmetric group $S_{N_c}$ the free theory of four Bosons and Fermions, at large-$N_c$~\footnote{$N_c$ is the number of fundamental strings in the F1-NS5 brane system. The NS5-brane charge is one.}.
Such interplay of symmetries and exact techniques makes this a prime playground to explore holography and discover new links between its underlying mathematical structures. In this letter I will first briefly review the integrable S~matrix for the $\AdS_3\times\S^3\times\T^4$ string sigma model, then specialize to the pure-NSNS case. There I shall highlight a new class of \textit{exceptional solutions} of the Bethe equations related to the long-string continuum~\cite{Maldacena:2000hw}. At $k=1$, I shall show that they yield a discrete  spectrum and match it to $\text{Sym}^{N_c}T^4$. 

\paragraph{Worldsheet integrability.}
I begin by summarising the main features of the integrable structure arising from the worldsheet theory in uniform light-cone gauge~\cite{Arutyunov:2004yx,Arutyunov:2005hd,Arutyunov:2006gs} for $\AdS_3\times \S^3\times \T^4$ strings, see also~\cite{Sfondrini:2014via}. The global symmetries $\psu(1,1|2)_{\sL}\oplus\psu(1,1|2)_{\sR}$ are broken by the gauge-fixing to an algebra containing \textit{half} of the supercharges, ($i,j=1,2$)
\begin{equation}
\label{eq:cealgebra}
\begin{aligned}
\acomm{\Ql{i}}{\Sl{j}}&=\delta^i_{\,j}\,\Hl\,,\qquad &\acomm{\Ql{i}}{\Qr{j}}&=\delta^i_{\,j}\,\C\,,\\
\acomm{\Qr{i}}{\Sr{j}}&=\delta^{j}_{\,i}\,\Hr\,,\qquad &\acomm{\Sl{i}}{\Sr{j}}&=\delta^j_{\,i}\,\C^\dagger.\\
\end{aligned}
\end{equation}
Here $\Hl$ is the combination of the $\su(1,1)_{\sL}$ Cartan element~$\mathbf{L}_{0}$ with the $\su(2)_{\sL}$ Cartan $\mathbf{J}^{3}$ that yields the $\psu(1,1|2)_{\sL}$ BPS bound: $\Hl\equiv\mathbf{L}_{0}-\mathbf{J}^{3}\geq0$. Similar formulae hold for the elements of $\psu(1,1|2)_{\sR}$ which I write with tildes. 
More interesting are the central extensions $\C, \C^\dagger$~\cite{Borsato:2012ud,Borsato:2013qpa} which couple the left and right generators and play a role similar to Beisert's central extension~\cite{Beisert:2005tm,Arutyunov:2006yd}. As it turns out~\cite{Borsato:2014hja}, on a state of worldsheet momentum~$p$ we have that $\C\approx\C^\dagger\approx h \sin (p/2)$, where $h$ is the amount of RR flux characterising the background. Importantly, $\C=\C^\dagger=0$ for physical states satisfying $p=0\,\text{mod}(2\pi)$ \textit{as well as for pure-NSNS backgrounds}.

\paragraph{Particle content.} 
All particles of the theory transform in short (\textit{i.e.}, supersymmetric) representations of the algebra~\eqref{eq:cealgebra} and contain two Bosons and two Fermions~\cite{Borsato:2013qpa,Borsato:2014hja}. The shortening condition reads~\cite{Borsato:2012ud}
\begin{equation}
\label{eq:shortening}
    \Hl\, \Hr = \C\,\C^\dagger\,.
\end{equation}
Let me introduce the \textit{total energy} $\Htot\equiv \Hl+\Hr$, as well as the orthogonal combination $\M\equiv \Hl-\Hr$. The former is the positive-semidefinite lightcone energy of a worldsheet particle, while the latter is the sum of $\AdS_3$ and $\S^3$ spins and is quantized. Then, evaluating  Eq.~\eqref{eq:shortening} on a representation yields the \textit{dispersion relation}~\cite{Hoare:2013lja,Lloyd:2014bsa}
\begin{equation}
\label{eq:dispersion}
    H(p) = \sqrt{\Big(\frac{k}{2\pi}p-m\Big)^2+4h^2\sin^2\frac{p}{2}}\,,\qquad m\in\mathbb{Z}\,.
\end{equation}
Here $h\geq0$ is the RR-flux strength, $k\in \mathbb{Z}_{\geq}$ is the NSNS-flux strength (\textit{i.e.}, the WZW level), while $p,m$ characterize each particle: $p\in\mathbb{R}$ is its worldsheet momentum, and $m\in\mathbb{Z}$ is the \textit{bound-state number} labelling each representation. More specifically: $m=0$ describes two representations related to $\T^4$ modes; $m=1$ gives one representation related to ``left'' $\AdS_{3}\times\S^3$ transverse modes and each $m\geq2$ describes bound states of left modes~\cite{Borsato:2013hoa}; similarly, for each $m=-1,-2, \dots$ there is one representation describing ``right'' modes or their bound states~\footnote{Unlike higher integrable AdS/CFT setups, particle and bound-state representations have identical dimension.}.
The zero-modes at $m=p=0$ signal symmetries: For Fermions, they yield a Clifford module of BPS states~\cite{Baggio:2017kza}. For Bosons, they seemingly yield an infinite degeneracy as one could add arbitrarily many of them without affecting energies. More precisely, they highlight the existence of a \textit{shift isometry} and (exactly as for the free Boson) they are not part of the spectrum, while the degeneracy is accounted for by an additional label --- the $\T^4$ momentum.
The knowledge of these representations is sufficient to construct the factorized S~matrix for generic $h,k$, see Refs.~\cite{Borsato:2014hja,Lloyd:2014bsa}. Henceforth I will consider pure-NSNS theories, \textit{i.e.}~$h=0$.

\paragraph{Pure-NSNS theories.}
At $h=0$ the theory becomes chiral: excitations move either to the left or to the right on the worldsheet at lightspeed $c=\pm k/2\pi$,
\begin{equation}
\label{eq:chiraldispersion}
    H(p)= \Big|\frac{k}{2\pi}p-m\Big|\,,\qquad
    m\in\mathbb{Z}\,.
\end{equation}
Let me call excitations with $c=+k/2\pi$ (respectively, $c=-k/2\pi$) \textit{chiral} (respectively, \textit{antichiral}).
The S~matrix is \textit{diagonal}
\begin{equation}
    \mathbf{S}(p_1,p_2)=e^{i \Phi(p_1,p_2)}\,\mathbf{1}\,,
\end{equation}
and the phase-delay takes the Dray-'t Hooft form~\cite{Dray:1984ha},
\begin{equation}
\label{eq:Smatrix}
    \Phi(p_1,p_2)=
\begin{cases}
   -\frac{k}{2\pi}p_1p_2\quad p_1\ \text{chiral},\ p_2\ \text{antichiral},\\
   +\frac{k}{2\pi}p_1p_2\quad p_2\ \text{chiral},\ p_1\ \text{antichiral},\\
   \ 0\qquad\qquad\text{else},
\end{cases}
\end{equation}
similarly to flat-space strings~\cite{Dubovsky:2012wk}. Owing to this, it is possible to solve the Bethe ansatz~\cite{Baggio:2018gct} as well as the mTBA~\cite{Dei:2018mfl} \textit{in closed form}. Let me briefly review the solution of the Bethe ansatz. Firstly, I introduce the notation~$\mu_i\equiv\mu(p_j)\equiv m_j \,\text{sgn}(\tfrac{k}{2\pi}p_j+m_j)$. One can then verify that (using subscript indices as a short-hand)
\begin{equation}
    \Phi_{jk}= \frac{1}{2}\big(p_j H_k - p_k H_j\big)-\frac{1}{2}\big(p_j \mu_k - p_k \mu_j\big)\,,
\end{equation}
and the Bethe ansatz reads $e^{i p_j R + i\sum_k \Phi_{jk}}=1$.
Recall now the level-matching condition $P\equiv\sum_k p_k=0$ $\text{mod}(2\pi)$. For simplicity, I take $P=0$ (the general case is analogous~\cite{Dei:2018mfl}):
\begin{equation}
\label{eq:logBY}
    p_j = \frac{2\pi n_j}{R_{\text{eff}}}\,,\quad 0\leq p_j<R_{\text{eff}}\,,\quad
    R_{\text{eff}}= R +\frac{H-\mu}{2}\,,
\end{equation}
where $H\equiv \sum_k H_k$ and $\mu=\sum_k \mu_k$. \footnote{To make the construction symmetric between chiral and antichiral modes, one can also consider the momentum interval $-R_{\text{eff}}<p_j<0$.}
On the other hand, using the dispersion~\eqref{eq:chiraldispersion} and $H\equiv \sum_k H_k$ I get
\begin{equation}
\label{eq:energyeq}
    H=\frac{k}{2\pi}\big(P^{(+)}-P^{(-)}\big)+\mu=\frac{k}{\pi}P^{(+)}+\mu\,,
\end{equation}
where I defined the sum of the momenta of the chiral (respectively, antichiral) excitations as $P^{(+)}$ (respectively, $P^{(-)}$) and used $P=P^{(+)}+P^{(-)}=0$. 
I introduce $N^{(+)}\equiv\sum_j^{\text{chiral}}n_j$ for the sum of chiral excitation numbers (and $N^{(-)}$ for antichiral ones).
Taking the sum of Eq.~\eqref{eq:logBY} and plugging it in Eq.~\eqref{eq:energyeq}, I obtain a quadratic equation in $H-\mu$ whose physical solution is
\begin{equation}
\label{eq:spectrumSC}
    H=\sqrt{R^2 +2 k (N^{(+)}+N^{(-)})}-R+\mu\,.
\end{equation}
As shown in Ref.~\cite{Dei:2018mfl}, Eq.~\eqref{eq:spectrumSC} remains correct \textit{even when all finite-size (``wrapping'') effects are included}, for instance by means of the mTBA.

\paragraph{New zero-energy excitations.}
From the dispersion \eqref{eq:chiraldispersion} it is in principle possible to have  excitations with $p_j\neq0$ but \textit{with zero energy}, $H_j=0$. This can only happen for special values of $R_{\text{eff}}$.  At the price of constraining~$R$, let me set $R_{\text{eff}}=W\in\mathbb{Z}_{>}$. Then Eq.~\eqref{eq:logBY} becomes
\begin{equation}
\label{eq:linearsol}
    p_j= \frac{2\pi n_j}{W}\,,\qquad n_j\in\{0,1,\dots, W-1\}\,.
\end{equation}
A zero-mode appears for $m_j=1$ if $n_j= W/k$, which may happen when $W=wk$ with $w\in\mathbb{Z}_{>}$. In this case, for any $k\geq2$ we encounter \textit{a novel zero-mode} to which we can associate \textit{a novel continuous momentum}~$t$. Let me emphasize that this momentum mode is entirely distinct from those related to the~$\T^4$ directions (which appear for $p=m=0$). The energy is then
\begin{equation}
\label{eq:spectrumqbigger1}
  k\geq2:\quad
    H = \frac{N^{(+)}+N^{(-)}}{w}+\mu+t^2,\qquad t\in \mathbb{R},
\end{equation}
where I added the energy due to $t$, supplemented by the level-matching condition
\begin{equation}
\label{eq:levelmatching}
    P=0\,\text{mod}(2\pi)\,,\qquad\text{\it i.e.}\ \
    N^{(+)}=N^{(-)}\,\text{mod}(kw)\,.
\end{equation}
I will argue below that~Eq.~\eqref{eq:spectrumqbigger1} describes the continuous representations (long strings) in the WZW model.

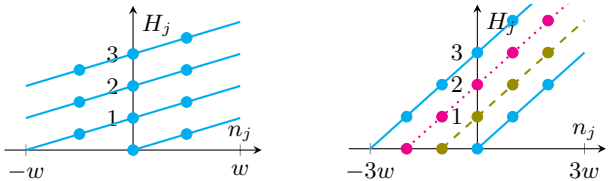
\begin{figure}
\centering

\begin{tikzpicture}
\begin{axis}[height=3.5cm, width=5cm,
xmin=-1.2,xmax=1.2,ymin=0,ymax=4.5, axis lines=center,xlabel=$n_j$,ylabel=$H_j$,ytick={0,1,2,3},xtick={-1,0,1},xticklabels={$-w$,$0$,$w$}]

\addplot[domain=0:1, color=cyan,thick]{x};
\addplot[domain=-1:1, color=cyan,thick]{x+1};
\addplot[domain=-1:1, color=cyan,thick]{x+2};
\addplot[domain=-1:1, color=cyan,thick]{x+3};

\addplot[color=cyan,mark=*, only marks] coordinates {
	(0,0)
	(0.5,0.5)
	(0,1)
	(0.5,1.5)
	(0,2)
	(0.5,2.5)
	(0,3)
	(0.5,3.5)
	(-0.5,0.5)
	(-0.5,1.5)
	(-0.5,2.5)
};

\end{axis}
\end{tikzpicture}%
\hspace*{1cm}%
\begin{tikzpicture}
\begin{axis}[height=3.5cm, width=5cm,
xmin=-1.2,xmax=1.2,ymin=0,ymax=4.5, axis lines=center,xlabel=$n_j$,ylabel=$H_j$,ytick={0,1,2,3},xtick={-1,0,1},xticklabels={$-3w$,$0$,$3w$}]

\addplot[domain=0:1, color=cyan,thick]{3*x};
\addplot[domain=-0.66:1, color=olive,thick, dashed]{3*x+1};
\addplot[domain=-1.32:1, color=magenta,thick, dotted]{3*x+2};
\addplot[domain=-1.98:1, color=cyan,thick]{3*x+3};

\addplot[color=cyan,mark=*, only marks] coordinates {
	(0,0)
	(0.33,1)
	(0.66,2)
};

\addplot[color=olive,mark=*, only marks] coordinates {
	(-0.33,0)
	(0,1)
	(0.33,2)
	(0.66,3)
};

\addplot[color=magenta,mark=*, only marks] coordinates {
	(-0.66,0)
	(-0.33,1)
	(0,2)
	(0.33,3)
	(0.66,4)
};

\addplot[color=cyan,mark=*, only marks] coordinates {
	(-0.66,1)
	(-0.33,2)
	(0,3)
	(0.33,4)
};

\end{axis}
\end{tikzpicture}
\caption{
The dispersion relation for different bound-state numbers ($m=0,1,2,3$) when momenta are quantized as in Eq.~\eqref{eq:linearsol}.
Left: at $k=1$ all energy levels are identified mod$w$.
Right: at $k=3$ there are three distinct trajectories mod$w$ and new zero-modes appear.
}
\label{fig:trajectories}
\end{figure}

\paragraph{The case $k=1$.}
Note that Eq.~\eqref{eq:spectrumqbigger1} fails at $k=1$. Indeed in that case it cannot be $n_j=w=W$ due to Eq.~\eqref{eq:linearsol}, and therefore no continuum arises. As the Bethe equations~\eqref{eq:linearsol} are essentially free, I can write a partition function. In Figure~\ref{fig:trajectories}, I highlighted another unique feature of~$k=1$: All bound-state energies fall on the same trajectory when we identify the mode numbers mod$w$. For the purpose of counting energy levels, bound-state representations effectively are equivalent to fundamental-particle representations with higher mode number --- somewhat like covering a length-$w$ circle with a line. Then the partition function for chiral modes at fixed~$w\in\mathbb{Z}_{>}$ is
\begin{equation}
\label{eq:Fw}
    F_w(q)=4\prod_{n=1}^{w-1}\frac{(1+q^{n/w})^4}{(1-q^{n/w})^4}
    \prod_{m=1}^\infty\prod_{n=0}^{w-1}\frac{(1+q^{m+n/w})^4}{(1-q^{m+n/w})^4},
\end{equation}
where $q$ is the chemical potential for~$\Hl$ and I excluded the Boson zero-modes (but kept the Fermion ones yielding a fourfold degeneracy). For antichiral modes and $\Hr$, I have $F_w(\tilde{q})$ so that in total the partition is given by $F_w(q)F_w(\tilde{q})$, subject to the level-matching constraint~\eqref{eq:levelmatching}. The complete spectrum arises when considering all $w\in\mathbb{Z}_{>}$.
In order to understand the physical meaning of this construction it is worth comparing with the insights gathered from CFT techniques.

\paragraph{WZW description.}
One may study the worldsheet theory as a supersymmetric $\sl(2)_{k}\oplus\su(2)_{k}$ WZW model (strictly speaking, for $k\geq2$) supplemented by free Fermions and free Bosons~\cite{Maldacena:2000hw,Maldacena:2000kv,Maldacena:2001km}. In that case, one finds that Eq.~\eqref{eq:spectrumSC} reproduces the states emerging from discrete representations of $\sl(2)_{k}$ provided that various quantities are opportunely identified~\cite{Dei:2018mfl}. States may come from discrete-series representations (possibly, spectrally flowed) or in spectrally-flowed continuous representations~\cite{Maldacena:2000hw}. Consider \textit{e.g.}\ a highest-weight state of the Ka\v{c}-Moody algebras with $\sl(2)$ charge $\ell_0$ and $\su(2)$ charge~$j_0$ in the NS sector, and act on that state with oscillators corresponding to the Ka\v{c}-Moody currents and to the free fields. The resulting state will be physical if the level-zero generator of the worldsheet Virasoro algebra annihilates it. The eigenvalue equation follows from the Sugawara construction: for a highest-weight representation,
\begin{equation}
\label{eq:WZWequation}
    -\frac{\ell_0(\ell_0-1)}{k}+
    \frac{j_0(j_0+1)}{k}+\mathscr{N}=0\,,
\end{equation}
where $\mathscr{N}$ is the collective (chiral) mode-number.
As $\mathscr{N}$ and $j_0$ are quantized, \eqref{eq:WZWequation} is an equation for~$\ell_0$. Knowing $\ell_0$, $\Hl$ follows, and the total~$\Htot$ comes from adding the antichiral sector and minding the level-matching condition, thereby reproducing~\eqref{eq:spectrumSC}. In particular this requires identifying~\cite{Dei:2018mfl}
\begin{equation}
    R_{\text{eff}}= \ell_0+j_0\,,\qquad
    N^{(+)}=\mathscr{N}\,,\quad
    N^{(-)}=\widetilde{\mathscr{N}}\,.
\end{equation}
For discrete representations, unitarity bounds restrict the allowed values of $\ell_0$ and $j_0$~\cite{Maldacena:2000hw}. For a ``spectrally flowed'' representation labelled by $w\in\mathbb{Z}_{>}$, one has that $\ell_0+j_0$ lies in infinitely many \textit{disconnected} intervals, 
\begin{equation}
    wk+\frac{1}{2}<\ell_0+j_0<
    (w+1)k-\frac{1}{2}\,.
\end{equation}
Therefore, $R_{\text{eff}}=wk$ lies \textit{outside} of the unitarity bounds for the discrete representations. However, precisely at these points we find the \textit{continuous representation}~\cite{Maldacena:2000hw}, which is consistent with the above appearance of Bosonic zero-modes, \textit{cf.}\ Eq.~\eqref{eq:spectrumqbigger1}.
Indeed the energy for continuous representations takes the form~\eqref{eq:spectrumqbigger1} up to a constant shift~\cite{Maldacena:2000hw,Eberhardt:2019qcl}.

\paragraph{Symmetric product orbifold CFT.}
The WZW description is plagued by unitarity issues at $k=1$, which require a more careful analysis~\cite{Giribet:2018ada,Gaberdiel:2018rqv, Eberhardt:2018ouy}. Following such an analysis it was argued that, precisely at $k=1$, the theory should become equivalent to a \textit{dual symmetric product orbifold} $\text{Sym}^{N_c}\T^4$.
Let me focus on this theory to show that it matches the partition function~\eqref{eq:Fw}. States are labeled by conjugacy classes of~$S_{N_c}$. I restrict to large $N_c$, as usual for the AdS/CFT spectral problem, so that conjugacy classes are in one-to-one correspondence with arbitrary products of disjoint cycles. I further restrict to \textit{single-cycle} states of length~$w\in\mathbb{Z}_{>}$. (Akin to single-trace operators in gauge theory, or single-particle states in supergravity.) I denote a $w$-cycle by $|\sigma_w\rangle$. Its energy depends on whether $w$ is even or odd~\cite{Gaberdiel:2018rqv}:
\begin{equation}
\label{eq:twistdimension}
    \mathbf{L}_0|\sigma_w\rangle=
    \widetilde{\mathbf{L}}_0|\sigma_w\rangle=\frac{w^2-\sin^2(\tfrac{\pi}{2}w)}{4w}|\sigma_w\rangle\,.
\end{equation}
Its $\su(2)$ charges vanish, $\mathbf{J}^3|\sigma_w\rangle=\widetilde{\mathbf{J}}^3|\sigma_w\rangle=0$.
On this state one acts with fractionally-moded Bosons $\alpha_{-n/w}^{iI}$, and Fermions $\psi^{aI}_{-n'/(2w)},\tilde{\psi}^{\tilde{a}I}_{-n'/(2w)}$, the latter carrying half-unit of $\su(2)_{\sL}$ and $\su(2)_{\sR}$ R-charge, respectively. Here $n'$ is odd or even depending on whether $w$ is odd or even, respectively.
Despite being the lightest state with respect to $\sl(2)$, $|\sigma_w\rangle$ is an excited state for $\Htot$ (except for the trivial $w=1$ cycle). It is convenient to construct a BPS reference state $|\Sigma_w\rangle$. This is obtained from $\sigma_w$ by filling a Fermi sea with  $\Htot\leq0$ Fermion-modes~\cite{Lunin:2001pw}. I briefly review this construction. For odd~$w$ I add precisely $w$ Fermion modes with $\textbf{J}^3$ charge $1/2$. Starting from the lightest, I add two modes (recall the index $I=1,2$) with $n'=1$, then two with $n'=3$, \textit{etc.}, until $n'=w-2$, for a total of $(w-1)$ oscillators, all with $\Hl<0$. Finally I add one single oscillator with $n'=w$ (and $\Hl=0$) obtaining precisely~$w$ oscillators; doing the same for antichiral oscillators I obtain a singlet under the auxiliary~$\su(2)_I$. For even~$w$, I pick the Ramond vacuum to have R-charge~$+1/2$; then I pick two oscillators with~$n'=2, 4,\dots, (w-2)$ and finally a single one with $n'=w$, for a grand total of $(w-1)$ oscillators. For either parity, using~\eqref{eq:twistdimension} I get
\begin{equation}
    \mathbf{J}^3|\Sigma_w\rangle
    =\widetilde{\mathbf{J}}^3|\Sigma_w\rangle
    =\frac{w}{2}|\Sigma_w\rangle\,,\quad
    \Htot|\Sigma_w\rangle=0\,,
\end{equation}
all other charges vanishing. This BPS state is the analogue of the $R_{\text{eff}}=w$ spin-chain vacuum  described above. Starting from $|\Sigma_w\rangle$, one may only act with oscillators which never decrease $\Hl$ or $\Hr$. For chiral excitations, there are four Bosons, with modes contributing $n_j/w$ to $\Hl$ with $n\geq1$; the Fermions are also four, and they too contribute $n_j/w$ with $n_j\geq1$ when both their R-charges and the shift in $n_j$ due to the Fermi sea are accounted for. Finally, a careful analysis of the Fermi sea shows that there are two Fermion zero-modes. This precisely yields $F_w(q)$ as in Eq.~\eqref{eq:Fw}. The constructions for antichiral excitations is analogous. Finally, one must impose the orbifold-invariance condition  $\sum_j n_j=\sum_j \tilde{n}_j\,\text{mod}\,w$, precisely reproducing the level-matching condition~\eqref{eq:levelmatching}.

\paragraph{Conclusions and outlook.}
I argued that the Bethe equations of Refs.~\cite{Baggio:2018gct,Dei:2018mfl} display a long-string continuum for $k\geq2$ and that at $k=1$ they yield a discrete spectrum, matching the single-cycle partition function of $\text{Sym}^{N_c}\T^4$. All these features rely on the absence of RR-fluxes: Even a tiny $h>0$ would spoil the chirality of Eq.~\eqref{eq:dispersion}. Indeed the long-string spectrum must disappear at $h>0$~\cite{Maldacena:2000hw,Eberhardt:2018vho} but little more is known about mixed-flux theories, as CFT techniques show limitations; I hope that this integrability treatment may, in time, prove helpful to understand them.
For $k\geq2$, Eq.~\eqref{eq:spectrumqbigger1} predicts a continuum but no gap from the BPS bound, unlike the CFT result~\cite{Maldacena:2000hw,Eberhardt:2019qcl}. The gap might arise from finite-volume effects related to the new zero modes, similar to Ref.~\cite{Dei:2018yth}. It would therefore be interesting to perform a full mTBA analysis at $k\geq2$.

Intriguingly, unlike for higher-dimensional AdS/CFT, this spectrum is described by algebraic equations. Technically, this is because wrapping may at most shift of the ground-state~\cite{Dei:2018jyj}. Physically, this suggests the existence of an underlying \textit{quantum-mechanical} description, whose Hilbert space and Hamiltonian would be extremely interesting to discover. It is tempting to imagine some spin-chain, but at least at $k\geq2$ the S~matrix~\eqref{eq:Smatrix} points to non-local interactions as discussed in Ref.~\cite{Marchetto:2019yyt}. Still, $k=1$ might be simpler.
The Hamiltonian would also the best tool to compute correlation functions (see Refs.~\cite{Eberhardt:2019ywk,Dei:2019osr,Dei:2020zui} for recent progress by CFT techniques), but possibly those might also  be tackled by \textit{hexagon tessellations}. (In AdS$_{5}$/CFT$_{4}$, these yield three-~\cite{Basso:2015zoa} and higher-point functions~\cite{Eden:2016xvg,Fleury:2016ykk} as well as non-planar ones~\cite{Eden:2017ozn,Bargheer:2017nne}.)
Finally, one could extend this analysis to $\AdS_{3}\times\S^{3}\times\S^{3}\times\S^{1}$ which possesses many of the features seen here, both from the CFT~\cite{Eberhardt:2019niq} and integrability viewpoint~\cite{Dei:2018jyj}.

\begin{acknowledgments}
\paragraph{Acknowledgments.}
I am grateful to Andrea Dei, Sibylle Driezen, Lorenz Eberhardt, Fiona Seibold and all the participants of the workshop \textit{Correlation Functions in Low-Dimensional AdS/CFT} held in Villa Garbald, Castasegna (CH) for stimulating discussions that lead to this letter. I would like to also thank Matthias Gaberdiel, Olof Ohlsson Sax and Bogdan Stefa\'nski jr.\ for useful related discussions. Special thanks to Andrea Dei, Lorenz Eberhardt and Sergey Frolov for providing feedback on a draft of this letter.
This work was  funded by ETH Zurich under Career Seed Grant No.~SEED-23-19-1 and by the Swiss National Science Foundation under Spark grant n. 190657.
\end{acknowledgments}

\end{document}